\documentclass[twocolumn,showpacs,preprintnumbers,amsmath,amssymb,superscriptaddress,prb]{revtex4}
\usepackage{epsfig}
\usepackage{graphicx}
\usepackage{color}

\begin{document}

\title{On the character of states near the Fermi level in (Ga,Mn)As: \\impurity to valence band crossover}
\author{T.~Jungwirth}
\affiliation{%
Institute of Physics ASCR v.v.i., Cukrovarnick\'a 10, 162 53 Praha
6, Czech Republic} \affiliation{School of Physics and Astronomy,
University of Nottingham,
  Nottingham NG7 2RD, United Kingdom}

\author{Jairo~Sinova}
\affiliation{Department of Physics, Texas A\&M University, College
Station, TX 77843-4242, USA}

\author{A.~H.~MacDonald}
\affiliation{Department of Physics, University of Texas at Austin,
Austin, Texas 78712-1081, USA}

\author{B.~L.~Gallagher}
\affiliation{School of Physics and Astronomy, University of
Nottingham, Nottingham NG7 2RD, United Kingdom}

\author{V.~Nov\'ak}
\affiliation{Institute of Physics ASCR v.v.i., Cukrovarnick\'a 10,
162 53 Praha 6, Czech Republic}

\author{K.~W.~Edmonds}
\affiliation{School of Physics and Astronomy, University of
Nottingham, Nottingham NG7 2RD, United Kingdom}

\author{A.~W.~Rushforth}
\affiliation{School of Physics and Astronomy, University of
Nottingham, Nottingham NG7 2RD, United Kingdom}

\author{R.~P.~Campion}
\affiliation{School of Physics and Astronomy, University of
Nottingham, Nottingham NG7 2RD, United Kingdom}

\author{C.~T.~Foxon}
\affiliation{School of Physics and Astronomy, University of
Nottingham, Nottingham NG7 2RD, United Kingdom}

\author{L.~Eaves}
\affiliation{School of Physics and Astronomy, University of
Nottingham, Nottingham NG7 2RD, United Kingdom}

\author{K.~Olejn\'{\i}k}
\affiliation{Institute of Physics ASCR v.v.i., Cukrovarnick\'a 10,
162 53 Praha 6, Czech Republic}

\author{J.~Ma\v{s}ek}
\affiliation{Institute of Physics ASCR v.v.i., Na Slovance 2, 182 21 Praha
8, Czech Republic}

\author{S.-R.~Eric~Yang}
\affiliation{Department of Physics, Korea University, Seoul 136-701, Korea}

\author{J.~Wunderlich}
\affiliation{Hitachi Cambridge Laboratory, Cambridge CB3 0HE, United Kingdom}

\author{C.~Gould}
\affiliation{Physikalisches Institut(EP 3), Universit\"{a}t W\"{u}rzburg, Am Hubland, 97074 W\"{u}rzburg, Germany}

\author{L.~W.~Molenkamp}
\affiliation{Physikalisches Institut(EP 3), Universit\"{a}t W\"{u}rzburg, Am Hubland, 97074 W\"{u}rzburg, Germany}

\author{T.~Dietl}
\affiliation{Institute of Physics, Polish Academy of Science, al. Lotnik´ow 32/46, PL 02-668 Warszawa, Poland}
\affiliation{Institute of Theoretical Physics, Warsaw University, PL 00-681 Warszawa, Poland}
\date{\today}

\author{H.~Ohno}
\affiliation{Laboratory for Nanoelectronics and Spintronics,
Research Institute of Electrical Communication, Tohoku University,
Katahira 2-1-1, Aoba-ku, Sendai 980-8577, Japan}
\affiliation{ERATO Semiconductor Spintronics Project, Japan Science and Technology Agency, Japan}

\begin{abstract}
We discuss the character of states near the Fermi level
in Mn doped GaAs, as revealed by a survey of dc transport and optical studies
over a wide range of Mn concentrations.  A thermally activated
valence band contribution to dc transport, a mid-infrared peak
at energy $\hbar\omega\approx 200$~meV in the ac-conductivity, and the hot photoluminescence
spectra indicate the presence of an impurity band in low doped ($\ll 1$\% Mn) insulating GaAs:Mn
materials.  Consistent with the implications of this picture, both the impurity band ionization energy inferred from the dc transport and the position of the mid-infrared peak move
to lower energies and the peak broadens with increasing Mn concentration.
In metallic materials with $>2$\% doping, no traces of Mn-related activated contribution can be identified in
dc-transport, suggesting that the impurity band has merged with the valence band. No discrepancies with
this perception are found when analyzing optical measurements in the high-doped GaAs:Mn.
A higher energy ($\hbar\omega\approx 250$~meV) mid-infrared feature which appears in the
metallic samples is  associated with
inter-valence band transitions. Its red-shift with increased doping
can be interpreted as a consequence of increased screening which narrows the localized-state valence-band tails
and weakens higher energy transition amplitudes.
Our examination of the dc and ac transport characteristics of GaAs:Mn is accompanied by comparisons with its
shallow acceptor counterparts, confirming the disordered valence band
picture of high-doped metallic GaAs:Mn material.
\end{abstract}

 \pacs{75.47.-m}


\maketitle

\section{Introduction}

GaAs is an intermediate band-gap III-V semiconductor, with $E_g=1.5$~eV at low temperatures,
in which an isolated Mn impurity substituting for Ga functions as an acceptor with an
impurity binding energy of intermediate strength,
$E_a^0=0.11$~eV.\cite{Chapman:1967_a,Blakemore:1973_a,Bhattacharjee:2000_a,Yakunin:2004_b,Madelung:2003_a}
Beyond a critical Mn concentration, Mn
doped GaAs exhibits a phase transition to a state in which the Mn
impurity levels overlap sufficiently strongly that the ground state
is metallic, {\em i.e.}, that states at the Fermi level are not bound
to a single or a group of Mn atoms but are delocalized across the
system.\cite{Matsukura:2002_a,Jungwirth:2006_a} In the metallic regime
Mn can, like a shallow acceptor (C, Be, Mg, Zn, e.g.),
provide delocalized holes with a low-temperature density
comparable to Mn density, $N_{\rm Mn}$.\cite{Ruzmetov:2004_a,MacDonald:2005_a,Jungwirth:2005_b}
($x=1$\% Mn-doping corresponds
to $N_{{\rm Mn}}=2.2\times 10^{20}$~cm$^{-3}$ in GaAs:Mn.)
The transition to the metallic state
occurs at larger doping in GaAs when doped with Mn than when doped
with shallow acceptors.  Experimentally the transition appears to
occur between $N_{{\rm Mn}}\approx 1\times 10^{20}$~cm$^{-3}$ and
$5\times 10^{20}$~cm$^{-3}$, as compared to the  $\sim 10^{18}$~cm$^{-3}$ critical
density for the shallow acceptors in GaAs.\cite{Silva:2004_a}

The hole binding potential of an
isolated substitutional Mn impurity is composed of long-range Coulomb,
short-ranged central-cell, and $sp-d$ kinetic-exchange potentials.\cite{Bhattacharjee:2000_a}
Because of the kinetic-exchange and central cell interactions, Mn acceptors are more
localized than shallow acceptors.  A crude estimate of the critical
metal-insulator transition density can be obtained with a
short-range potential model, using the experimental binding energy
and assuming an effective mass of valence band holes, $m^{\ast}=0.5m_e$.  This model implies an isolated
acceptor level with effective Bohr radius $a_0=(\hbar^2/2m^{\ast}E_a^0)^{1/2}=10\,\AA$.
The radius $a_0$ then equals the Mn impurity spacing scale $N_{{\rm Mn}}^{-1/3}$
at $N_{{\rm Mn}}\approx 10^{21}$~cm$^{-3}$. This explains qualitatively the higher
metal-insulator-transition critical density in Mn doped GaAs compared to
the case of systems doped with shallow, more hydrogenic-like acceptors
which have binding energies $E_a^0\approx 30$~meV.\cite{Madelung:2003_a,Silva:2004_a}
\begin{figure}[h]

\hspace*{0cm}\includegraphics[width=0.9\columnwidth,angle=0]{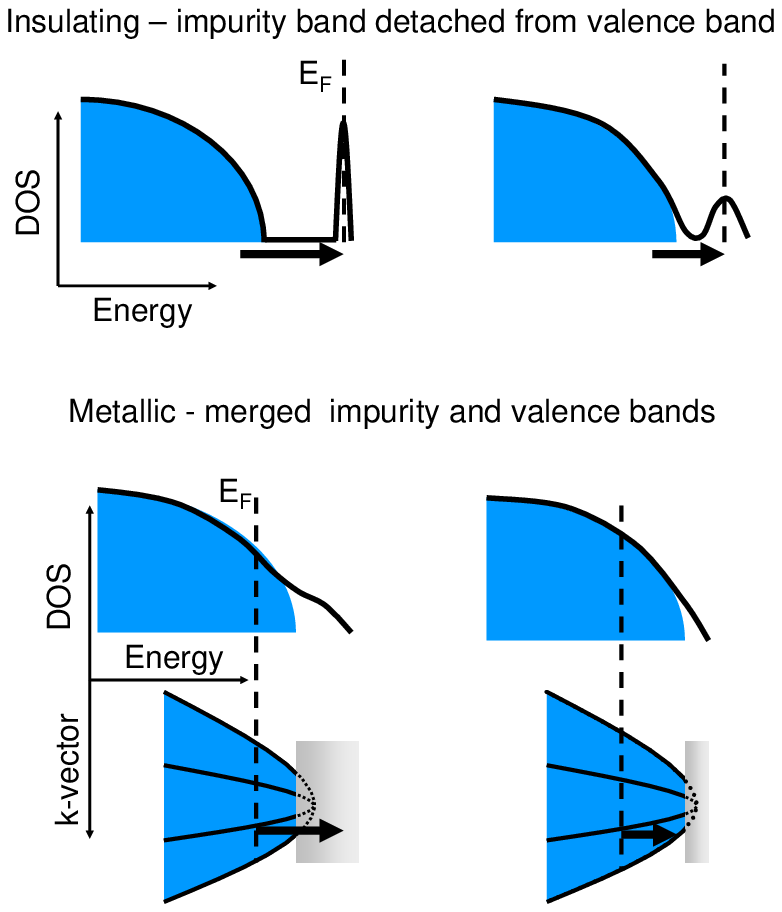}

%
\caption{Schematic illustrations of the impurity band regime for low doped insulating
GaAs:Mn (upper panels) and of the disordered valence band regime for the high doped metallic
GaAs:Mn (lower panels). Splitting of the bands in the ferromagnetic state is omitted for
simplicity. For each regime the cartoons are arranged from left to right according
to increasing doping. Blue
areas indicate delocalized states, white and grey areas localized states. Arrows highlight the red shift in the impurity band ionization
energy and the red shift in the inter-valence-band transitions, respectively, which are discussed
in Sections~II to IV.
}
\label{Figure1}
\end{figure}

The intermediate and magnetic character of extrinsic GaAs:Mn makes the physics
near its metal-insulator transition even more complex than in the
shallow, non-magnetic  acceptor counterparts and difficult to describe quantitatively.
However, as it is often the case in semiconductors, important phenomena
occur at or near
this transition doping region. In GaAs:Mn, the most
remarkable among these is the onset of
carrier mediated ferromagnetism. The understanding of magnetic properties
of these doped semiconductors which are close to the metal-insulator transition can therefore only
emerge from the
understanding of
the spectroscopic nature of carriers in this regime. 

Unlike the metal-insulator phase transition, which is sharply defined
in terms of the temperature $T=0$ limit of the conductivity, the crossover
in the character of states near the Fermi level in semiconductors with increased doping is
gradual.\cite{Shcklovskii:1984_a,Lee:1985_a,Paalanen:1991_a,Jungwirth:2006_a,Dietl:2007_b}
At very weak doping, the Fermi level resides inside
a narrow  impurity band (assuming some compensation) separated from the valence band by an energy gap of a magnitude
close to the impurity binding energy.
In this regime strong electronic correlations are
an essential element of the physics and a single-particle picture has limited
utility.  Well into the metallic state, on the other hand, the impurities are sufficiently
close together, and the various potentials which contribute to the binding
energy of an isolated impurity are sufficiently screened, that
the system is best viewed as an imperfect crystal with disorder-broadened and
shifted host Bloch bands.
In this regime, electronic correlations are usually less strong
and a single-particle picture often suffices.

Although neither picture is very helpful for describing the physics in the crossover regime
which spans some finite
range of dopings, the notion of the impurity band on the lower doping side from the crossover
and of the disordered host band on the higher doping side from
the crossover  still have a clear qualitative meaning. The former implies that
 there is a deep minimum in the density-of-states
between separate impurity and host band states.
In the latter case the impurity band and the host band merge into one inseparable
band whose tail may still contain localized states depending on the carrier concentration and disorder.
We note that terms overlapping and merging impurity and valence bands describe the same
basic physics in GaAs:Mn. This is because
the Mn-acceptor states span several unit cells even in the very dilute
limit and  many unit cells as the impurity band broadens with increasing doping.
The localized and the delocalized
Bloch states then have a similarly
mixed As-Ga-Mn {\em spd}-character. This applies to
systems on either side of the metal-insulator transition. We also point out that
for randomly distributed Mn dopants in GaAs the impurity band
is not properly modeled by assuming a Bloch-like, effective mass character.

The impurity band and the disordered valence band regimes
are schematically illustrated in Fig.~\ref{Figure1}.
In Sections~II to IV of this article we discuss a
survey of previously published dc and ac transport data and new dc measurements of GaAs:Mn
materials grown in the Nottingham\cite{Campion:2003_a} and Prague\cite{Kopecky:2006_a}
molecular-beam-epitaxy (MBE) systems
spanning a wide range of Mn concentrations.  Our comprehensive study complements
previous investigations of specific narrower doping regions. It shows
that the picture of the crossover from the impurity band at low dopings
to the disordered valence band at high dopings, illustrated
in  Fig.~\ref{Figure1} and generally accepted for the shallow acceptor counterparts
to GaAs:Mn, applies
well to the intermediate extrinsic GaAs:Mn. The perception of merged impurity and valence bands
in highly Mn-doped metallic GaAs is now also firmly established in the microscopic theory community\cite{Krstajic:2004_a,Hwang:2005_a,Wierzbowska:2004_a,Jungwirth:2006_a,Yildirim:2006_a,Popescu:2007_a,Dietl:2007_b} and provides a qualitative, and often a semiquantitative description of micromagnetic and magnetotransport
characteristics of bulk and microstructured (Ga,Mn)As ferromagnets.\cite{Matsukura:2002_a,Jungwirth:2006_a,Dietl:2007_b}

Our article, which we believe further establishes the valence band nature
of the Fermi level states in metallic GaAs:Mn, is
timely as the topic is still not fully settled. Several
recent interpretations of optical measurements in
GaAs:Mn \cite{Burch:2006_a,Sapega:2005_a,Sapega:2006_a}
have favored the impurity band picture of
highly doped metallic materials. These works provided additional motivation for the analysis presented  here,
for which we conclude that the optical data in the high-doped GaAs:Mn
are not inconsistent with the disordered valence band nature
of states near the Fermi energy, inferred from dc-transport.

Finally, before moving to the technical sections  let us remark that comparisons between samples with
different nominal Mn concentrations have to be treated with caution. The Mn doping parameter should
only be considered as an approximate guideline for qualitative discussions of doping trends.
This applies namely to epitaxial materials with
higher Mn concentrations in which nominal, growth-rate controlled Mn doping parameter
can deviate significantly from
the amount of Mn impurities incorporated in the acceptor-like Ga-substitutional positions. Moreover,
the partial densities of various types of Mn impurities and of other compensating
and structural defects can vary significantly depending on  applied growth conditions.

\section{Low-doped insulating ${\rm\bf GaAs:Mn}$ systems with impurity bands}

\begin{figure}[ht]
\vspace*{.5cm}
\hspace*{-0cm}\includegraphics[width=1.0\columnwidth,angle=-0]{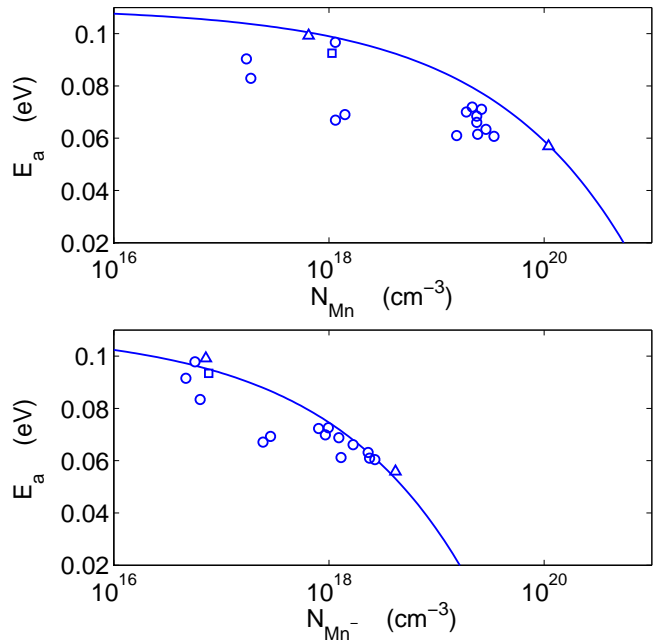}

\vspace*{-0cm} \caption{Mn doping dependent energy gap between the impurity
band and valence band measured in the bulk equilibrium grown
GaAs:Mn and plotted as a function of total Mn density (top panel) and ionized
Mn denisty (bottom panel). All data are replotted from Ref.~\protect\onlinecite{Blakemore:1973_a}}
\label{Figure2}
\end{figure}

\begin{figure}[ht]
\vspace*{-0cm}
\includegraphics[width=1.0\columnwidth,angle=-0]{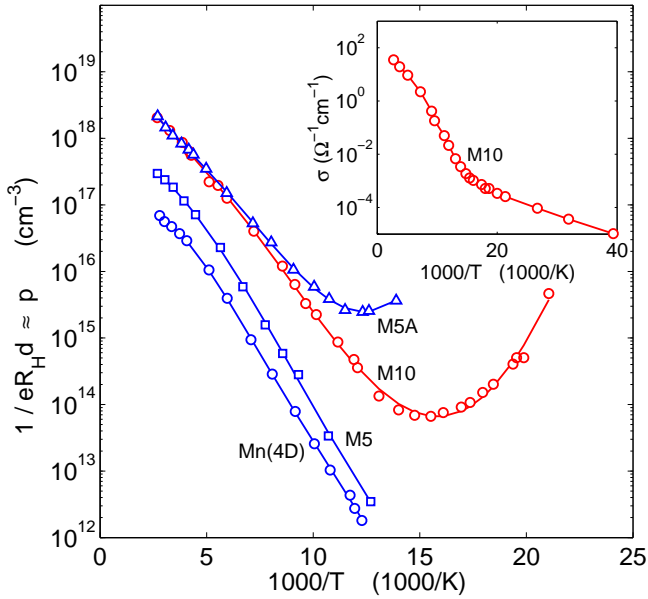}

\vspace*{-0cm} \caption{Hole densities determined from the Hall coefficient measurement as
a function of temperature in bulk equilibrium grown GaAs:Mn. Inset: temperature-dependent
longitudinal conductivity. From
Refs.~\protect\onlinecite{Brown:1972_a,Blakemore:1973_a}. Sample Mn(4D) has Mn
density
$1.6\times10^{17}$~cm$^{-3}$, M5: $1.1\times10^{18}$~cm$^{-3}$,
M10: $1.9\times10^{19}$~cm$^{-3}$, Mn5A: $9.3\times10^{19}$~cm$^{-3}$.}
\label{Figure3}
\end{figure}

\begin{figure}[ht]
\vspace*{-0cm}

\hspace*{-0cm}\includegraphics[height=0.9\columnwidth,angle=-0]{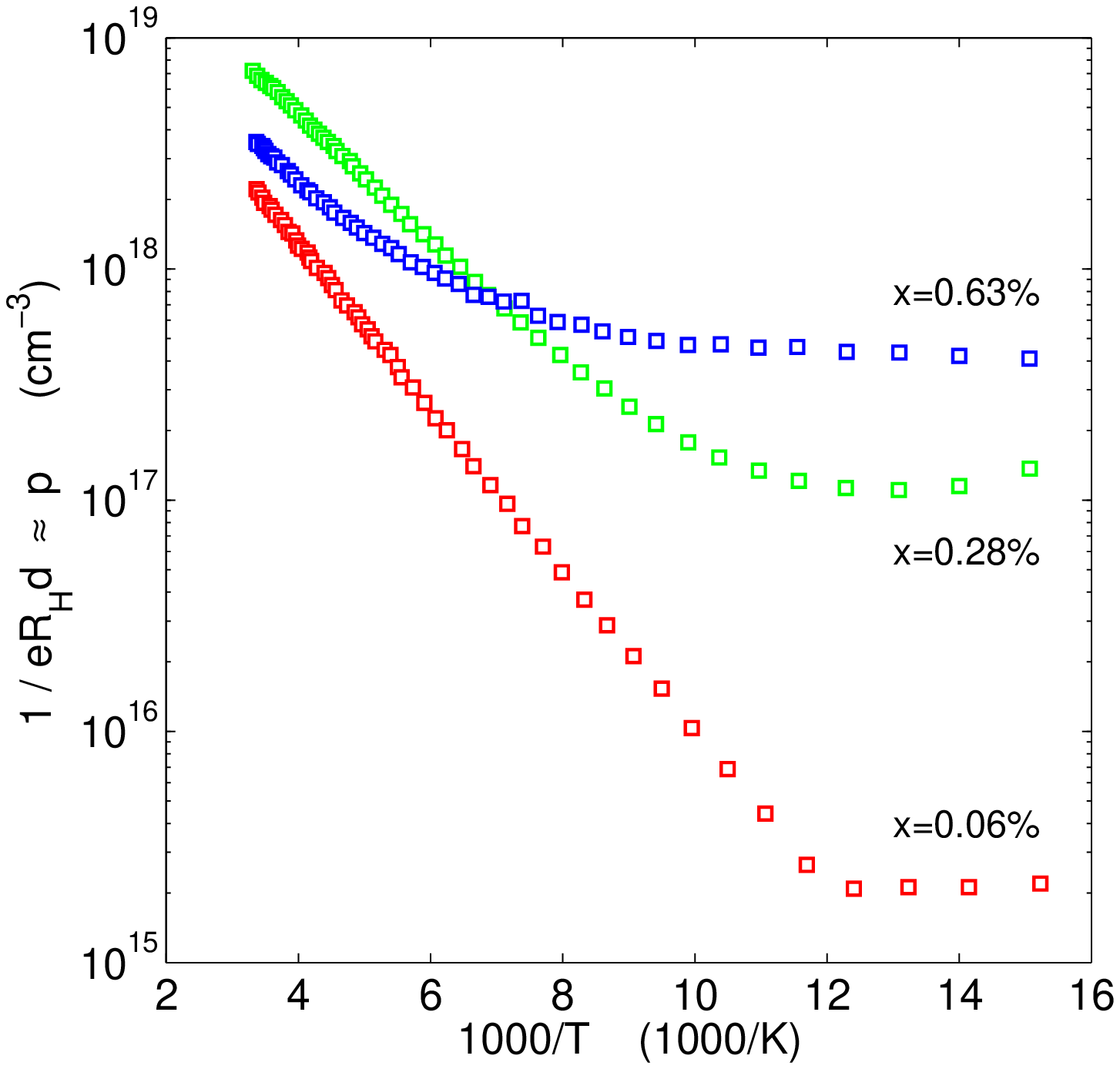}

\vspace*{-0cm} \caption{Same as Fig.~\protect\ref{Figure3}
for MBE grown (at 400$^{\circ}$C) GaAs:Mn. From
Ref.~\protect\onlinecite{Poggio:2005_a}.}
\label{Figure4}
\end{figure}

Narrow impurity bands are expected and have been clearly observed in Mn doped GaAs samples with
carrier densities much lower than the metal-insulator transition density,
for example in equilibrium grown bulk
materials with $N_{\rm Mn}=10^{17}-10^{19}$~cm$^{-3}$.\cite{Brown:1972_a,Woodbury:1973_a,Blakemore:1973_a}
The energy gap between the impurity band and the valence band, $E_a$, can be measured
by studying the temperature dependence of longitudinal and Hall conductivities, which show activated
behavior because of thermal excitation of holes from the
impurity band to the much more conductive valence band.\cite{Blakemore:1973_a,Woodbury:1973_a,Marder:1999_a}
Examples of these measurements are shown in
Figs.~\ref{Figure2}-\ref{Figure4}.

The activation energy decreases with increasing
Mn density, following roughly the form\cite{Blakemore:1973_a}
\begin{equation}
\label{eq:ea_insulator}
E_a=E_a^0[1-(N_{\rm Mn}/N^c_{\rm Mn})^{1/3}].
\end{equation}

The lowering of impurity binding energies at larger $N_{Mn}$, which is expected to
scale with the mean impurity
separation as expressed in Eq.~(\ref{eq:ea_insulator}), is apparent already in the equilibrium grown
bulk materials with $N_{\rm Mn}=10^{17}-10^{19}$~cm$^{-3}$.  The trend continues in epitaxially grown (at 400$^{\circ}$C)
GaAs:Mn with $N_{\rm Mn}\approx 1\times 10^{19}$ and $6\times
10^{19}$~cm$^{-3}$.\cite{Poggio:2005_a}
When this trend is extrapolated using Eq.~(\ref{eq:ea_insulator})
it places the disappearance of the gap at Mn dopings of $\sim 1$\%.
Note that this estimate has a large scatter depending also on whether all Mn or
only ionized Mn impurities are considered when fitting the data by Eq.~(\ref{eq:ea_insulator})
(see Fig.~\ref{Figure2}).\cite{Blakemore:1973_a}

For $N_{\rm Mn}\approx 1\times 10^{19}$~cm$^{-3}$ ($x\approx 0.06$\%),
$E_a\approx 70$~meV and a sharp crossover from impurity band hopping
conduction to activated valence band conduction is observed at $T\approx 80-100$~K (see Figs.~\ref{Figure3}
and \ref{Figure4}).
Near the crossover, $N_{\rm Mn}/p\sim10^{3}-10^{5}$ where $p$ is the valence-band hole density, consistent with the much
higher mobility of valence band holes compared to impurity band holes in this
regime for which the distinction is clearly still valid and useful.
(Note that in the bulk equilibrium-grown samples
the temperature at which the crossover occurs appears to be
very sensitive to the detailed disorder configuration at this Mn doping level,
with some samples being strongly insulating and others showing
signatures of filamentary conducting channels.\cite{Brown:1972_a,Woodbury:1973_a,Poggio:2005_a})

For the $N_{\rm Mn}\approx 6\times 10^{19}$~cm$^{-3}$ ($x\approx 0.3$\%) epitaxial
sample,\cite{Poggio:2005_a} $E_a\approx 50$~meV and the crossover from impurity band
conduction to activated valence band conduction can still be identified
at $T\approx 140$~K (see Fig.~\ref{Figure4}).
The notion of the very low mobility impurity band separated
from a Bloch valence band starts to blur at this doping as the $N_{\rm Mn}/p$ ratio near the
conduction-type crossover is only about 100.

\begin{figure}[ht]
\vspace*{-0cm}
\hspace*{-0cm}\includegraphics[width=1.0\columnwidth,angle=-0]{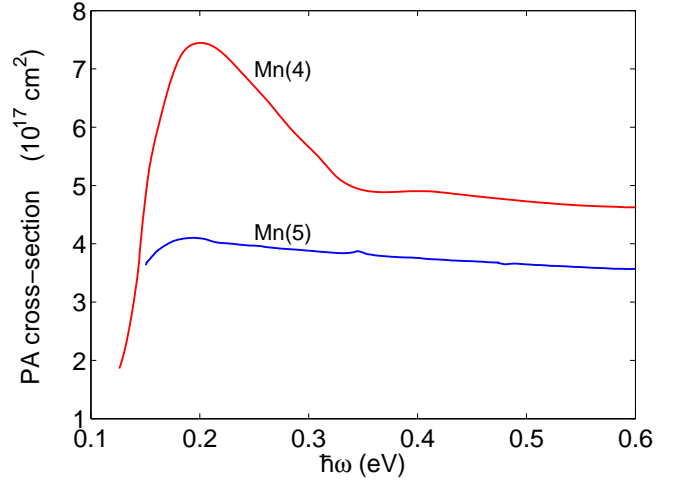}
\vspace*{-0cm} \caption{Infrared photoabsorption crossection measurements in bulk
GaAs:Mn. From Ref.~\protect\onlinecite{Brown:1972_a}. Mn(4):
$1.7\times10^{17}$~cm$^{-3}$, Mn(5): $9.3\times10^{18}$~cm$^{-3}$.
}
\label{Figure5}
\end{figure}

In the dilute limit where the system has a narrow insulating impurity band,
the ac conductivity is expected to show a broad mid-infrared feature due to
impurity band to valence band transitions, peaked at
energy between $E_a$ and $2E_a$.\cite{Brown:1972_a,Anderson:1975_a,Fleurov:1982_a} Lower energy
peak positions correspond to shallow acceptors with
weakly bound impurity states composed of a narrow range of valence band wavevector Fourier
components. The peak near $2E_a$ and a line shape $\sigma(\hbar\omega)\sim
E_a^{1/2}(\bar\omega-E_a)^{3/2}/(\hbar\omega)^3$ correspond to a short-range impurity potential for which a larger
interval of valence band states contributes to the transition amplitude. Consistent with these
expectations
a peak at $\hbar\omega\approx 200$~meV is
observed in the weakly doped ($N_{\rm Mn}\sim 10^{17}$~cm $^{-3}$)
GaAs:Mn samples (see Fig.~\ref{Figure5}).\cite{Chapman:1967_a,Brown:1972_a,Singley:2002_a}  At $N_{\rm
Mn}\sim 10^{19}$~cm $^{-3}$, the peak is slightly red shifted
($\hbar\omega\approx 180$~meV) and significantly broadened.\cite{Brown:1972_a} Both
the red-shifting and the broadening are expected consequences of
increasing overlap between acceptor levels.

In the dilute doping disorder free limit, hot photoluminescence (HPL)
can provide another optical spectroscopy tool, when combined with a reliable measure of the heavy hole kinetic energy at the valence band to conduction band excitation energy, to measure the position of a narrow
impurity band with respect to the top of the valence band.\cite{Twardowski:1985_a} For a
GaAs:Mn sample doped with $\sim 10^{17}$~cm$^{-3}$
of Mn, a HPL spectrum has been obtained\cite{Sapega:2005_a} (see also Fig.~\ref{Figure9})
that is consistent with excitations from the valence band to the conduction band
followed by transitions of the hot electrons (before their energy is dissipated in the
conduction band) into the impurity level.
An edge  can be identified in the spectrum which corresponds to
an impurity level $\sim 120\pm 10$~meV above the top of the valence band, in agreement with
the above dc and infrared ac transport data and other experiments.\cite{Chapman:1967_a,Blakemore:1973_a,Bhattacharjee:2000_a,Yakunin:2004_b,Madelung:2003_a}

We conclude that in the low doping regime, the temperature-dependent dc transport
data and the ac conductivity data consistently indicate the presence of an
impurity band and the decrease in $E_a$ as doping brings the impurity band
closer to the valence band. Note that the activated dc transport marker of the impurity band is
common to both the intermediate Mn acceptor and the shallow acceptors in GaAs.\cite{Silva:2004_a}
The infrared ac conductivity feature, on the other hand, appears to be  a more subtle and less
reliable tool for identifying the presence of the impurity band. To our knowledge, there are
no reports of the impurity
band transitions in infrared ac measurements for GaAs doped with shallow acceptors. In these systems,
even when weakly doped and still insulating, the reported infrared ac conductivity peaks are ascribed
to inter-valence-band transitions due to non-zero thermally activated or metallic hole densities
in the valence band.\cite{Braunstein:1962_a,Huberman:1991_a,Songprakob:2002_a} The inter-valence-band
transitions in metallic GaAs:Mn will be analyzed in detail in Section~IV.

\section{Impurity-band to disordered-valence-band crossover and high-doped metallic ${\rm\bf GaAs:Mn}$}

\begin{figure}[ht]
\vspace*{-.3cm}

\hspace*{-0cm}\includegraphics[width=.93\columnwidth,angle=-0]{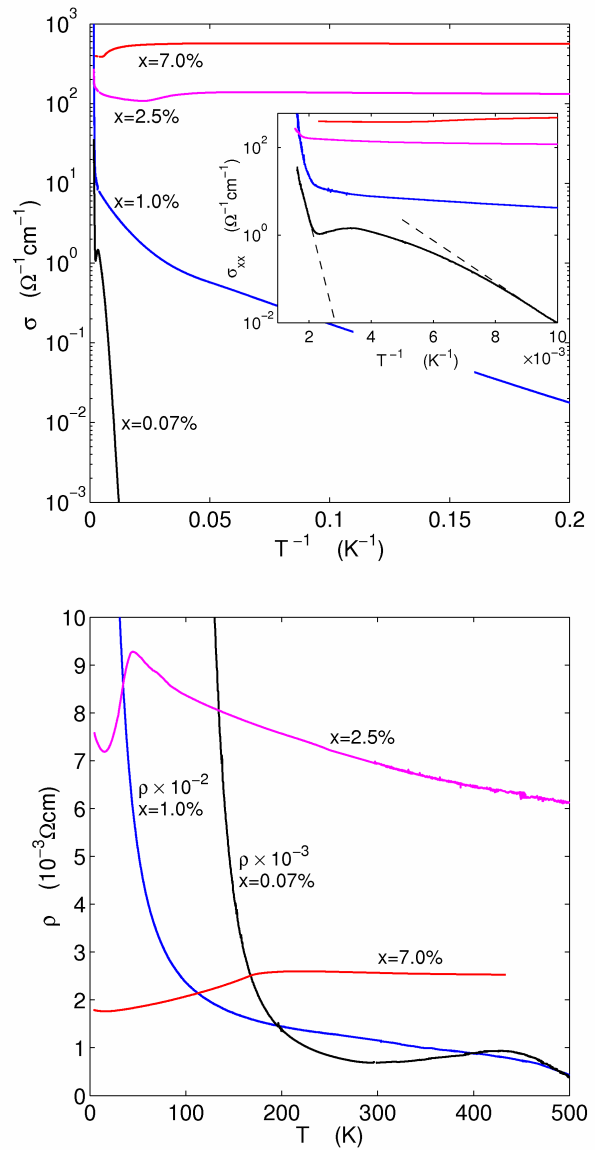}
\vspace*{-0.0cm}

\caption{Comparison of longitudinal conductivities (upper panel) and resistivities (lower panel)
of 0.07\% and 1\% doped insulating paramagnetic GaAs:Mn, and 2.5\% and 7\% Mn doped ferromagnetic
($T_c$ approximately 40~K and 170~K, resp.)
metallic GaAs:Mn grown
in the Prague MBE system with As$_4$ flux. The 0.07\% doped
material was grown at 530$^{\circ}$, the other materials were
grown by low-temperature MBE (240-200$^{\circ}$).
The 0.07\% doped sample shows clear GaAs:Mn impurity band -- valence band activation at low temperatures,
impurity band exhaustion, and the onset of activated transport over GaAs band gap
at high temperatures. The 2.5\% and 7\% doped samples show only weak conductance variations associated with the onset of ferromagnetism until the activated transport over GaAs band gap takes over.}
\label{Figure6}
\end{figure}

At higher Mn doping around $\approx 1$\%, the conductivity curves
can no longer  be separated into
an impurity-band dominated contribution at low temperatures and an activated valence-band dominated
contribution at higher temperatures, even though the samples are still insulators (see Fig.~\ref{Figure6}).
In metallic materials with $x>2$\%, no signatures of Mn-related hole
activation is observed when temperature is swept to values as high
as 500~K, {\em i.e.} to $T\approx E_a^0/2$, and before the intrinsic conductance
due to activation across the GaAs band-gap takes over (see Fig.~\ref{Figure6}).

\begin{figure}[ht]
\vspace*{-0cm}

\hspace*{-0cm}\includegraphics[width=0.9\columnwidth,angle=-0]{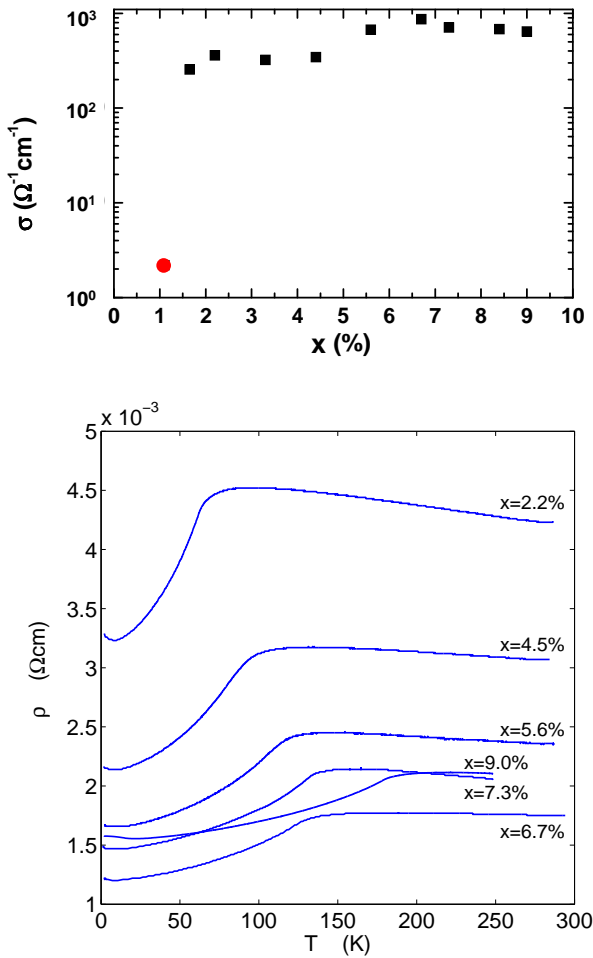}
\vspace*{-0cm} \caption{dc transport measurements in
optimally annealed, metallic ferromagnetic GaAs:Mn ($T_c$'s between approximately
50~K and 170~K) grown by the Nottingham low-temperature MBE
system with As$_2$ flux. Upper panel: conductivities at 10~K plotted as a function of Mn doping;
insulating 1\% doped material grown under the same conditions
is included for comparison (red dot). Lower panel:
temperature dependence of longitudinal resistivities.}
\label{Figure7}
\end{figure}

For completeness, we show in Fig.~\ref{Figure7} temperature-dependent resistivity
curves for
a set of metallic, annealed high conductivity ferromagnetic samples, which, within experimental uncertainty of ~20\%,
show no compensation and conductivities up to
$\approx 900$~$\Omega^{-1}{\rm cm}^{-1}$. We note here that the high-doped metallic GaAs:Mn samples
show only weak
temperature-dependence of the conductivity associated with the onset of ferromagnetism and
that the conductivity varies slowly with Mn composition in the metallic samples but changes dramatically in going from 1 to 1.5\% Mn in these samples.
We also note that no marked dependence of the hole density on temperature is observed for
metallic samples.\cite{Ruzmetov:2004_a}

\begin{figure}[ht]
\vspace*{-0cm}
\hspace*{-0cm}\includegraphics[width=.7\columnwidth,angle=-90]{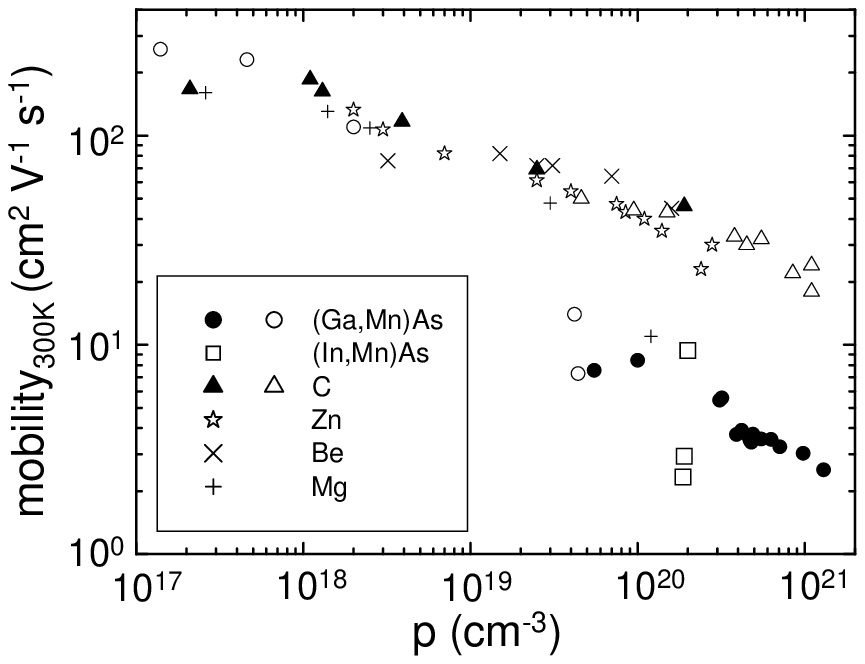}
\vspace*{-0cm} \caption{Room temperature Hall mobilities as a function hole concentration
for GaAs doped with Mn (filled circles - present results, open circles - Ref.~\protect\onlinecite{Moriya:2003_a}),
C (filled triangles - present results, open triangles - Ref.~\protect\onlinecite{Yamada:1989_a},
Zn (stars - Ref.~\protect\onlinecite{Glew:1984_a}), Be (X - Ref.~\protect\onlinecite{Parsons:1983_a}),
Mg (plus - Ref.~\protect\onlinecite{Kim:2001_a}), as well as InAs doped
with Mn (square - Ref.~\protect\onlinecite{Schallenberg:2006_a,Lee:2007_b}).
Hole concentrations were obtained
from low temperature high field Hall measurements for the
ferromagnetic
metallic GaAs:Mn and InAs:Mn films, and from room temperature Hall
measurements for the other films.
}
\label{Figure8}
\end{figure}

The absence of the impurity band
in high-doped GaAs:Mn is further evident by comparisons with other related materials
(see Fig.~\ref{Figure8}) for which the disordered valence band picture has not been questioned. The
narrower-gap InAs:Mn counterparts to GaAs:Mn,
{\em e.g}, in which Mn acts as a shallower acceptor have similarly low magnitudes and similar temperature-dependence
of the conductivity.\cite{Schallenberg:2006_a,Lee:2007_b,Ohya:2003_a} Comparably low conductivities (or mobilities) are also found
in epitaxially grown GaAs doped with $\sim 10^{20}$~cm$^{-3}$ of the shallow non-magnetic acceptor
Mg.\cite{Kim:2001_a} Note that the highest mobilities in p-type GaAs with the doping levels $\sim 10^{20}-10^{21}$~cm$^{-3}$
are achieved with Zn and C and these are only $5-10\times$ larger as compared to
GaAs:Mn (see Fig.~\ref{Figure8}).\cite{Glew:1984_a,Yamada:1989_a}.

We also remark that
valence band calculations treating disorder
in the Born approximation overestimate the experimental dc conductivities
of metallic GaAs:Mn by less than
a factor of ten\cite{Jungwirth:2002_c,Sinova:2002_a} and that any sizable discrepancy is  removed by
exact-diagonalization calculations\cite{Yang:2003_b} (see also Fig.~\ref{Figure12}) which account for
strong disorder and localization effects. This provides
another confirmation for the plausibility of the disordered valence band picture.

\begin{figure}[ht]
\vspace*{0.cm}
\hspace*{0.cm}\includegraphics[width=0.75\columnwidth,angle=-90]{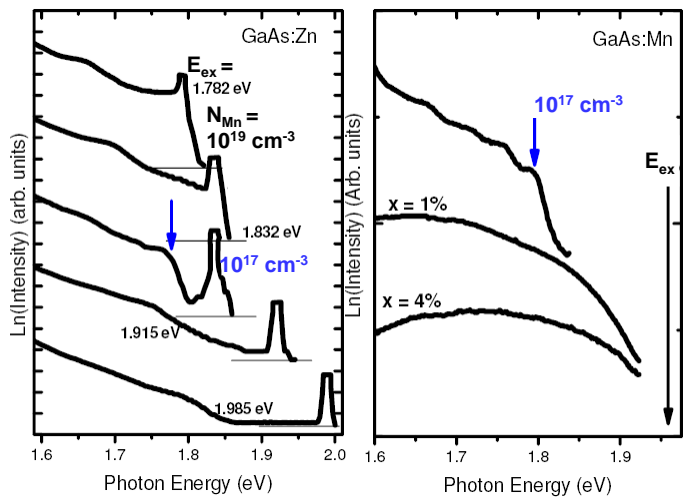}
\vspace*{0cm} \caption{Left panel: HPL spectra of GaAs doped with  $\sim 10^{17}$~cm$^{-3}$ of Zn (see blue arrow) and
$\sim 10^{19}$~cm$^{-3}$ of Zn. From Ref.~\onlinecite{Twardowski:1985_a} (replotted in logarithmic scale).
 Right panel:
HPL spectra of GaAs doped with  $\sim 10^{17}$~cm$^{-3}$ of Mn (see blue arrow), and of $\approx 1$\% and $\approx 4$\%
GaAs:Mn materials.
From Ref.~\onlinecite{Sapega:2005_a}.
$E_{ex}$ is the HPL excitation energy. Arrows indicate the spectral feature corresponding
to impurity band transitions in the low-doped
materials.
}
\label{Figure9}
\end{figure}

Similarities between GaAs doped with Mn and with shallow acceptors, Zn in particular, are
also found in the experimental
HPL spectra, as shown in Fig.~\ref{Figure9}.\cite{Twardowski:1985_a,Sapega:2005_a} While
for doping levels  $\sim 10^{17}$~cm$^{-3}$  a sharp
onset of the luminescence at frequencies corresponding to transitions into the impurity bands in
these materials
is clearly detected, the HPL
impurity band marker is not apparent at high dopings.
For the metallic GaAs:Zn with $N_{\rm Zn}\approx
10^{19}$~cm$^{-3}$ and for the 4\% doped GaAs:Mn,
the spectra show only broad luminescence with no clear indication of an onset
but rather a featureless decrease continuing out towards the laser energy.

The first infrared ac conductivity measurements in  high-doped GaAs:Mn,
reported in Refs.~\onlinecite{Nagai:2001_a,Hirakawa:2002_a}, were inconclusive.
Only two samples were compared in both of these works, with
approximately 3\% and 5\% nominal Mn-doping. The absence of a clear mid-infared peak in the 5\% material
in Ref.~\onlinecite{Nagai:2001_a} and the re-entrant insulating behavior of the higher doped sample in
Ref.~\onlinecite{Hirakawa:2002_a} suggest that the growth at this early stage had not been optimized so as
to minimize the formation of unintentional defects and to achieve appreciable growth reproducibility.

\begin{figure}[ht]
\vspace*{-0.1cm}

\hspace*{0cm}\includegraphics[height=2.0\columnwidth,angle=0]{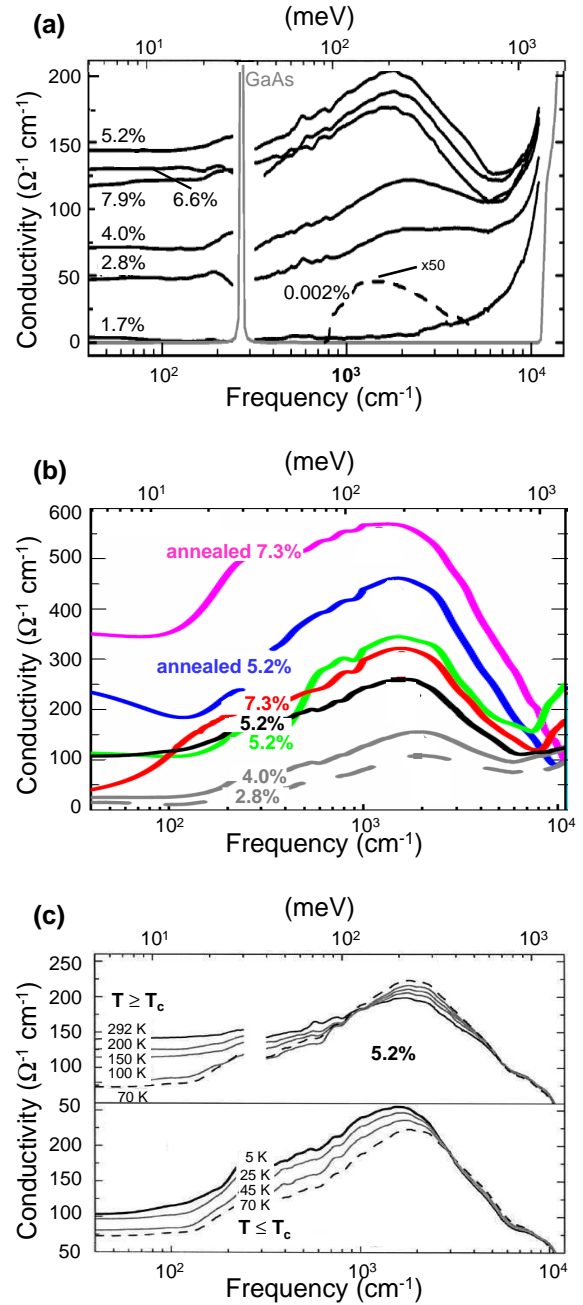}

\vspace*{-0cm} \caption{Infrared absorption measurements of GaAs:Mn. Top panel: comparison of GaAs, low-doped
GaAs:Mn and high-doped as-grown GaAs:Mn materials at 292~K. Middle panel: Comparison of high-doped as-grown and
annealed GaAs:Mn samples at 7~K. Bottom panel: Absorption measurements for the 5.2\%
as-grown GaAs:Mn below and above the ferromagnetic transition temperature.
From Refs.~\protect\onlinecite{Singley:2002_a,Singley:2003_a,Burch:2006_a}}
\label{Figure10}
\end{figure}

The more recent and more systematic
infrared absorption measurements\cite{Singley:2002_a,Singley:2003_a,Burch:2006_a}
for an insulating sample close to the metal-insulator transition and for a set
of metallic highly doped GaAs:Mn materials are shown in Fig.~\ref{Figure10}.
The ac conductivity of the $1.7$\% doped insulating sample is featureless in the mid-infrared
region.\cite{Singley:2002_a,Singley:2003_a}
This can be interpreted as a natural continuation of the trend, started in the low doped samples, in which
the impurity band moves closer to the valence band and broadens with increasing doping,
becoming eventually undetectable in the ac conductivity spectra. Note that the disappearance
of the impurity band transitions can in principle result also
from a strong compensation.\cite{Singley:2003_a} A strong
unintentional compensation in epitaxial GaAs:Mn at the $\sim 1$\% doping level is not typical, however.
The 1.7\% doped sample from Ref.~\onlinecite{Singley:2003_a} has a
room temperature conductivity in the dc limit of about 5~$\Omega^{-1}{\rm cm}^{-1}$, which presumably
drops to a much lower value at 5~K.\cite{Singley:2003_a} This is consistent
with the $\sigma(T)$ dependence of the insulating 1\% sample from the series of epitaxial materials in
Fig.~\ref{Figure6} which show no signatures of strong compensation over the entire range of doping
from 0.07-7\%.

A new mid-infrared feature emerges in the metallic $>2\%$ doped samples at frequencies,
$\hbar\omega\approx 250$~meV, and is then red-shifted by $\approx
80$~meV as the doping is further increased to approximately 7\% without showing a marked
broadening (see
Fig.~\ref{Figure10}).\cite{Singley:2002_a,Burch:2006_a}
The association of this peak to an impurity band is implausible because of (i)
the absence of the activated dc-transport counterpart in the high-doped metallic samples,
(ii) the blue-shift of this  mid-infrared feature with respect to
the impurity band transition peak in the $N_{\rm
Mn}\sim 10^{19}$~cm $^{-3}$ sample, (iii) the appearance of the peak at frequency
above $2E_a^0$, and (iv) the absence of a marked broadening of the peak
with increased doping.
In the following section we argue that this peak in
the infrared conductivity of metallic GaAs:Mn is
consistent with  the picture of merged impurity and valence bands in these materials.

\section{Red shift of the mid-infrared absorption above insulator-to-metal transition}
\begin{figure}[ht]
\vspace*{1.5 cm}

\hspace*{-0cm}\includegraphics[width=.9\columnwidth,angle=-90]{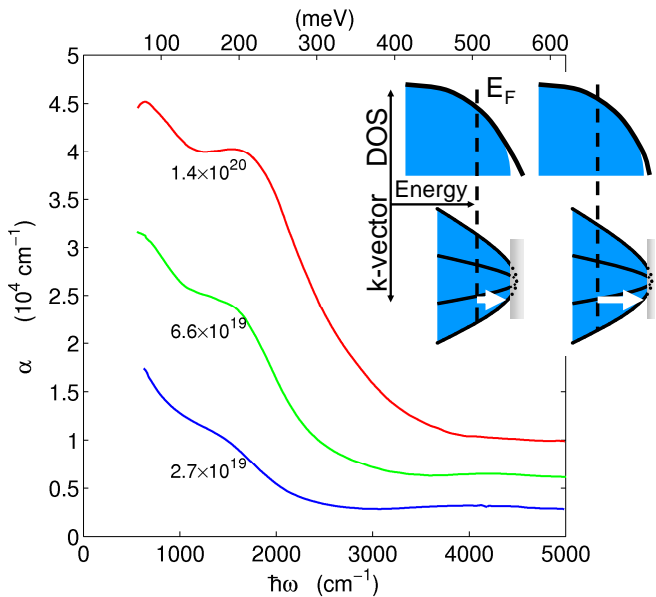}
%
\vspace*{-0cm}
 \caption{Infrared absorption measurements of metallic GaAs:C. From
Ref.~\protect\onlinecite{Songprakob:2002_a}. Inset illustrates the origin of the blue shift
of the mid-infrared peak in these highly  metallic systems.}
\label{Figure11}
\end{figure}

A mid-infrared peak has been observed in  metallic GaAs heavily doped ($\sim
10^{19}-10^{20}$~cm$^{-3}$) with shallow hydrogenic carbon acceptors and explained in terms
of inter-valence-band transitions between heavy holes and
light holes.\cite{Songprakob:2002_a} This peak blue-shifts with
increasing doping as the
Fermi energy moves further into the valence and, consequently,
the transitions move to higher energies (see Fig.~\ref{Figure11}).\cite{Songprakob:2002_a,Sinova:2002_a}
The conclusions of
previous sections suggest that the mid-infrared peaks in the
metallic GaAs:Mn materials should be of a qualitatively similar origin.
The red-shift seen
in GaAs:Mn can be understood by recalling that while GaAs doped with
$\sim 10^{19}-10^{20}$~cm$^{-3}$ C is already far on the metallic
side, GaAs doped with $\sim 10^{20}-10^{21}$~cm$^{-3}$ of Mn is
still close to the metal-insulator transition.

The merging of the impurity band and the valence band near the
transition is illustrated schematically in Fig.~\ref{Figure1}.
For metallic GaAs:Mn close to the metal-insulator transition,
states in the band tail
can still be expected to remain localized and spread out in energy
(see lower panels in Fig.~\ref{Figure1}).
The red shift of the inter-valence-band transitions can result from an
increased metallicity with increasing doping.
At lower doping the abundance of localized states in the broad
valence band tail enables  transitions which take
spectral weight from the low frequency region and
provide a channel for higher-energy transitions.
As doping increases, the valence band tail narrows
because of increased screening.  The
inter-valence-band absorption peak then red-shifts as
the low-frequency  part of the spectrum adds spectral weight.

\begin{figure}[ht]
\vspace*{-0cm}

\hspace*{-0cm}\includegraphics[width=.7\columnwidth,angle=-90]{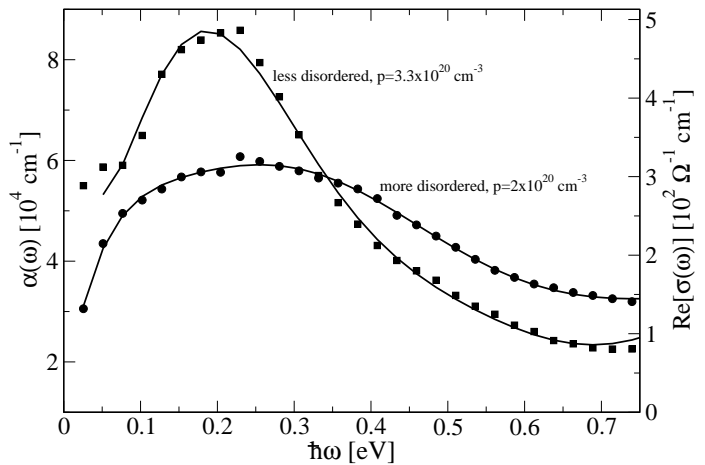}
\vspace*{-0cm} \caption{Exact diagonalization study showing the red shift
of the valence band mid-infrared peak for the Kohn-Luttinger
model assuming lower hole concentration and stronger disorder due to larger number
of compensating defetcs (dots)
and higher hole density and weaker disorder (rectangles). Lines are
10th-order polynomial fits to the numerical data to guide the eye.}
\label{Figure12}
\end{figure}

A support for this picture can be found in  finite-size
exact-diagonalization calculations reported in Ref.~\onlinecite{Yang:2003_b}
which were intended to address the
influence of strong disorder on the ac conductivity in metallic GaAs:Mn close to the metal
insulator transition.  In these numerical data, shown in Fig.~\ref{Figure12},
the mid-infrared feature is shifted to lower energies in the higher hole
density more conductive system.
In both experimental and numerical calculation cases,
the red-shift of the peaks is accompanied by an increase of the dc conductivity.
In experiment, this correlation is observed
both by comparing different samples and, in a given sample,
over the entire range of temperatures that was studied, below and above
the ferromagnetic transition temperature
(see Fig.~\ref{Figure10}).\cite{Singley:2002_a,Singley:2003_a,Burch:2006_a}

As the metal-insulator transition is approached from the metal side in a
multi-band system, it appears that the spectral weight that is lost from the disappearing
Drude peak, centered on zero frequency, is shifted to much higher energies, perhaps due to
transitions to localized valence band tail states as suggested by the
cartoon in Fig.\ref{Figure1}.  There is a useful sum rule for the conductivity integrated over
the energy range corresponding to the semiconductor band-gap,
provided that the Fermi energy is not too deep in the
valence band and that the valence band tails do not slide too deeply
into the gap.\cite{Sinova:2002_a,Yang:2003_b}
The sum-rule optical masses derived on the basis
of the ac conductivity  calculations for GaAs:Mn with the Fermi level in the
valence band are
consistent with experiment to within  a factor of two to
three.\cite{Sinova:2002_a,Yang:2003_b,Burch:2006_a}  (A more quantitative
comparison is hindered by
inaccuracies in the theoretical modeling
of the GaAs:Mn valence bands and
uncertainties associated with the high-frequency
cut-off imposed on the experimental sum rules and with the determination of material
parameters, hole densities in particular.)

We emphasize that no general practical sum rule exists for the low-frequency
part of the ac conductivity and that
a Drude fit of the spectra near the dc limit would tend to strongly overestimate band
effective masses as the metal-insulator transition is approached.
In GaAs:Mn in particular, most of the experimental ac conductivity curves in Fig.~\ref{Figure10} as well as
the numerical simulations in Fig.~\ref{Figure12}
do not show a maximum at zero frequency\cite{Singley:2002_a,Singley:2003_a,Burch:2006_a,Yang:2003_b}
which makes the Drude fitting questionable.

Finally we point out that the absolute conductance values of the mid-infrared conductivity peak
found experimentally in metallic GaAs:Mn materials, which are expected to be relatively insensitive to
disorder, agree with the values
predicted theoretically by the inter-valence-band transition
calculations.\cite{Sinova:2002_a,Yang:2003_b}
This includes both the  calculations treating disorder
within the Born approximation\cite{Sinova:2002_a}, which overestimate the dc conductivity, and
the exact-diagonalization studies shown in Fig.~\ref{Figure12} which account for
strong disorder and localization effects.\cite{Yang:2003_b}

\section{Summary}
In summary, a number of dc transport and optical measurements have been
performed during the past four decades to elucidate the nature of states
near the Fermi level in GaAs:Mn in the insulating
and metallic regimes. A detail examination of these studies, complemented with new
dc transport data and comparisons with shallow acceptor counterparts to GaAs:Mn, show
that  impurity band markers
are consistently seen in the insulating low-doped materials. Similar consistency
is found when analyzing experiments in the high-doped metallic materials, in this
case strongly favoring
the disordered valence band picture.
Our conclusions support the established theoretical description of
GaAs:Mn based on a variety of different {\em ab initio} and
phenomenological approaches
which explain
many experimental magnetic and magnetotransport properties of these dilute moment ferromagnetic
semiconductors.

{\em Acknowledgements. }
  We acknowledge fruitful and stimulating discussions with Dimitri Basov, Kenneth Burch, Elbio Dagotto, Alexander Finkelstein, Konstantin Kikoin, Yuri Kusrayev,
  Adriana Moreo, Victor Sapega, and Carsten Timm. This work was supported by ONR under Grant No.\ ONR-N000140610122, by the
  NSF under Grants No.\ DMR-0547875, SWAN-NRI, by the SRC-NRI (SWAN), by EU Grant NANOSPIN
  IST-015728, by EPSRC Grant GR/S81407/01, by GACR and AVCR Grants
  202/05/0575, FON/06/E002, AV0Z1010052, and LC510. J. Sinova is a Cottrell
  Scholar of Research Corporation. S.-R. Yang acknowledges support from the
  Second Brain 21 Project.

\end{document}